ms.tex toplevelfile
\documentclass{IEEEtran}
\pdfoutput=1 
\usepackage{mathtools}
\usepackage{amsmath,graphicx}
\usepackage{xcolor}
\usepackage{xspace}
\usepackage{url}

\makeatletter
\def\@copyrightspace{\relax}
\makeatother

\newcommand{\celsius}{$^\circ$C }

\begin{document}\sloppy

\title{Is Leakage Power a Linear Function of Temperature?}
%

\author{\IEEEauthorblockN{Hameedah Sultan}
\IEEEauthorblockA{School of Information Technology\\
Indian Institute of Technology Delhi\\
Email: hameedah@cse.iitd.ac.in}\\
\and
\IEEEauthorblockN{Shashank Varshney}
\IEEEauthorblockA{VLSI Design Tools and Technology\\
Indian Institute of Technology Delhi\\
Email: shashank.jvl17@ee.iitd.ac.in}\\
\and
\IEEEauthorblockN{Smruti R. Sarangi}
\IEEEauthorblockA{Computer Science and Engineering\\
Joint Professor in Electrical Engineering\\
Indian Institute of Technology Delhi\\
Email: srsarangi@cse.iitd.ac.in}}

\maketitle

%
\begin{abstract}
In this work, we present a study of the leakage power modeling techniques commonly used in the architecture community. We further provide an analysis of the error in leakage power estimation using the various modeling techniques. We strongly believe that this study will help researchers determine an appropriate leakage model to use in their work, based on the desired modeling accuracy and speed.
\end{abstract}

\section{Introduction}
Leakage power is now a significant part of modern sub-micron devices. 
Subthreshold leakage power dominates the total leakage power and is also very strongly dependent on temperature. As a result, several areas of architectural design and optimization require a leakage model. The requirements of all such areas vary immensely. For instance, in the early stages of the design process, the modeling accuracy is not of much importance, however the modeling speed is of great importance. Due to an insufficient awareness of the less complex leakage models possible, researchers often skip modeling leakage power completely, or use overly simplistic assumptions. 
As a result, we present a study of the various leakage models commonly used in the architecture community, along with the speed and accuracy of each model. 
Our goal is to enable researchers determine an appropriate leakage model that meets their requirements. 

\section{Methodology}
\label{sec:method}

In this section, we first present the leakage power obtained using the BSIM 4 equation. SPICE simulation results for 28 nm bulk CMOS technology are presented next, followed by SPICE simulation results for a 7nm FinFET using predictive models. Next we present a review of the prevalent leakage models, and analyse these models.

\subsection{BSIM Equation}
The standard subthreshold leakage power is given by the simplified BSIM 4 equation \cite{bsim}:

\begin{equation}
\label{eqn:bsim}
 P_{leak} \propto v_{T}^{2} * e^{\frac{V_{GS}-V_{th}-V_{off}}
{\eta * v_{T}}} (1-e^{\frac{-V_{DS}}{v_{T}}})
\end{equation}
where, $v_T$ is the thermal voltage ($kT/q$), $V_{th}$ is the threshold voltage, $V_{off}$ is the offset voltage in the
sub-threshold region and $\eta$ is a constant.

Equation~\ref{eqn:bsim} clearly shows that the leakage power is exponentially dependent on temperature. However, modeling an exponential relationship is very complex; hence several approximations to the BSIM equation have been proposed to reduce the modeling complexity.

We too used the BSIM equation to model leakage power within the standard  operating range of real ICs (40\celsius to 80\celsius), using the set of parameters listed in Table~\ref{tab:bsimparam}. We also used a linear approximation to find out the error obtained with respect to the speed up gained.
The value of the constant of proportionality has been set
such that the leakage power at ambient temperature (45\celsius) is $0.1~W$.
Next we fit a linear model (regression line) to the results obtained. 

\begin{figure}[!htbp]
\centering
\includegraphics[scale = .4,trim = {0cm 6cm 0cm 6cm}, clip=true]{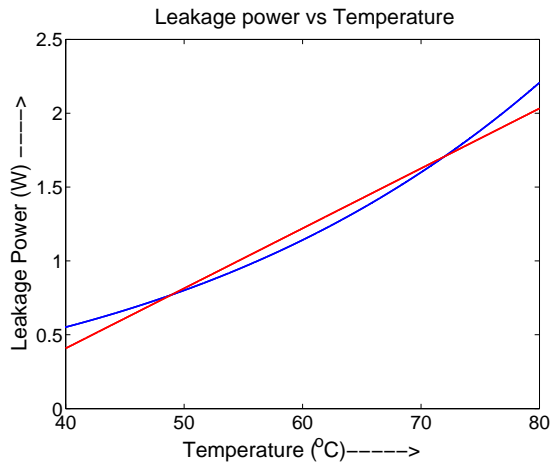}
\caption{Leakage power vs temperature using the BSIM equation\label{fig:bsim}}
\end{figure}

\begin{figure}[!htbp]
\centering
\includegraphics[scale = .4,trim = {0cm 6cm 0cm 6cm}, clip=true]{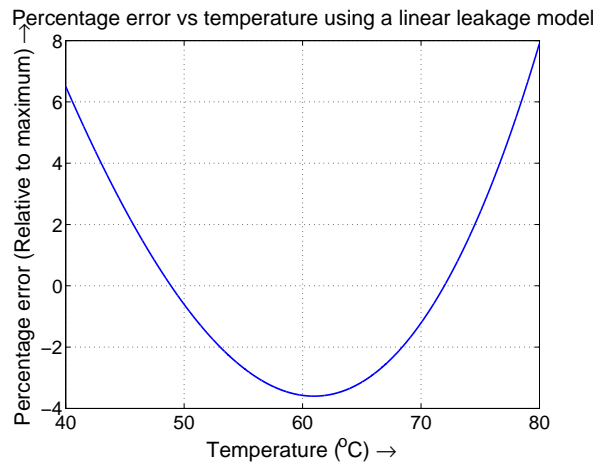}
\caption{Percentage error with linear leakage using the BSIM equation\label{fig:bsimlinerror}}
\end{figure}

\begin{table}[!htb]
\small
 \begin{center}
\caption{Parameters used for calculating leakage using the BSIM equation\label{tab:bsimparam}} \begin{tabular}{|p{2cm}|l|} 
\hline
Parameter & Value \\
\hline
$T_{amb}$& 318.15 K \\
\hline
$V_{GS} = V_{DS}$ & 0.7 V \\
\hline
$V_{th}$ & $0.15 - 0.004*(T - Tamb)~V$\\
\hline
$V_{off}$ & 0.0024 V \\
\hline
$\eta$ & 2 \\
\hline
$P_{leak0}$ & 0.1 W\\
\hline
\end{tabular}
\end{center}
\end{table}
.
The leakage power obtained using the BSIM equation, along with those with a linear approximation is shown in Figure~\ref{fig:bsim}. The percentage error obtained using the linear approximation is shown in Figure~\ref{fig:bsimlinerror}.

Table~\ref{tab:bsimspeed} lists the computation time obtained for a linear leakage model, a quadratic model and the complete BSIM equation, for a temperature values between 40 and 80\celsius, with increments of 0.1\celsius. 
\begin{table}[!htb]
\small
 \begin{center}
\caption{ Computation time for the BSIM model and its approximations\label{tab:bsimspeed}} \begin{tabular}{|p{4cm}|l|} 
\hline
Model & Computation Time (us) \\
\hline
BSIM Equation & 74.18\\
\hline
Linear approximation & 18.73\\
\hline
Quadratic approximation & 21.95 \\
\hline
\end{tabular}
\end{center}
\end{table}

Thus we conclude that a linear leakage model introduces an error of less than 5\%, and maybe sufficient for the early stages of the design process, where the accuracy is not of primary concern.

\subsection{SPICE Simulation Results}
\begin{figure}[!htbp]
\centering
\includegraphics[scale = .4,trim = {0cm 6cm 0cm 6cm}, clip=true]{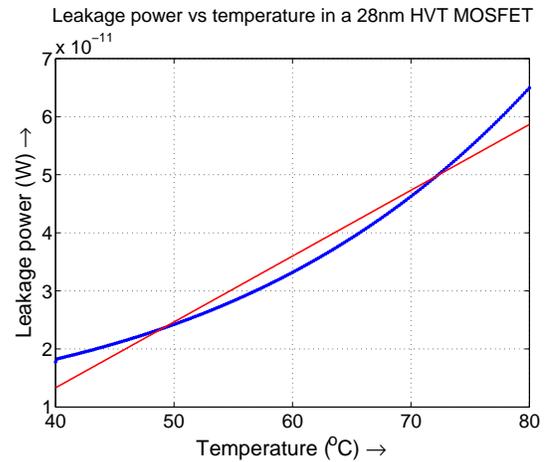}
\caption{Leakage power vs temperature for a 28 nm HVT MOSFET\label{fig:28hvtPT}}
\end{figure}

\begin{figure}[!htbp]
\centering
\includegraphics[scale = .4,trim = {0cm 6cm 0cm 6cm}, clip=true]{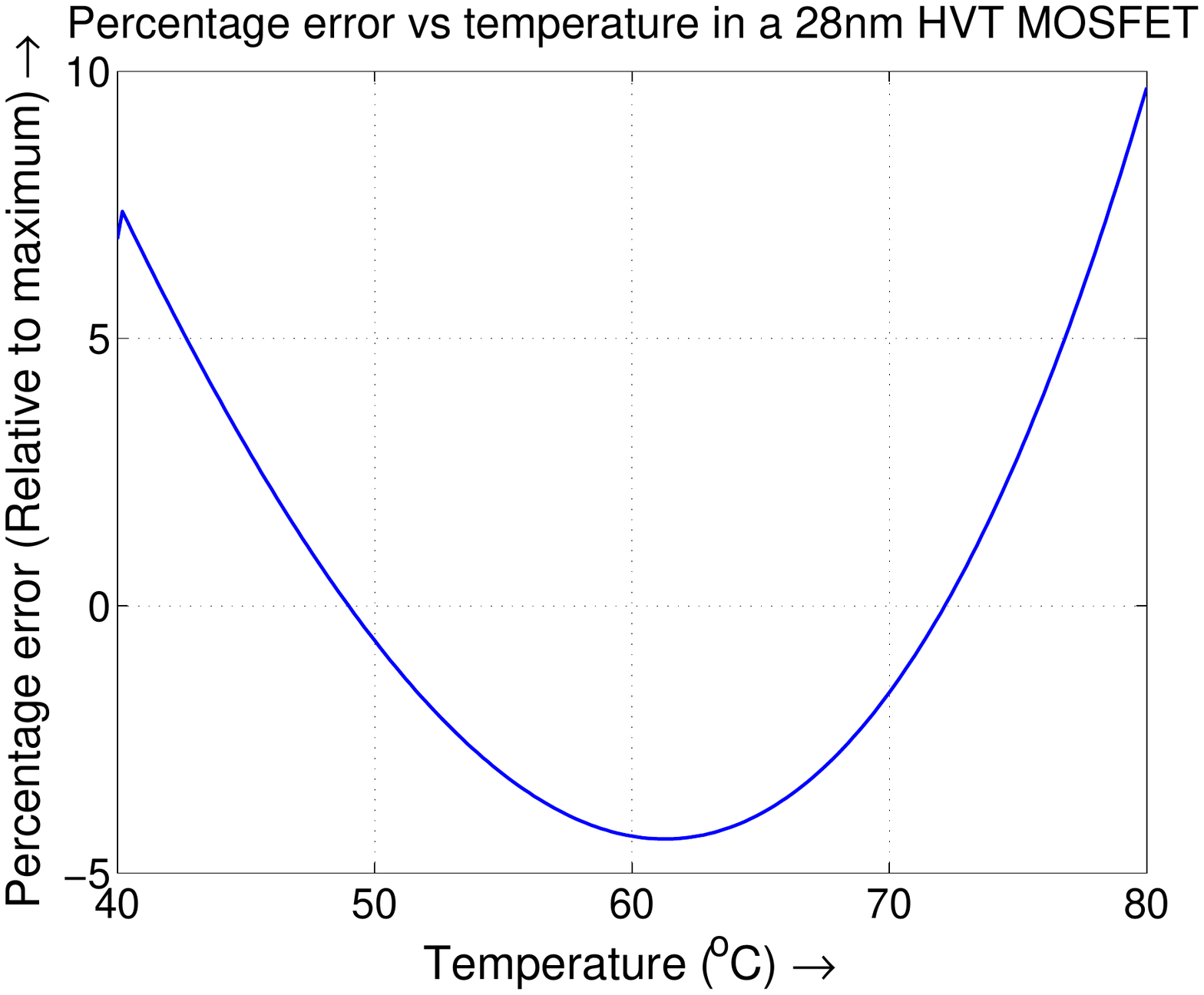}
\caption{Percentage error with a linear leakage model for a 28 nm HVT MOSFET\label{fig:hvterror}}
\end{figure}

\begin{figure}[!htbp]
\centering
\includegraphics[scale = .4,trim = {0cm 6cm 0cm 6cm}, clip=true]{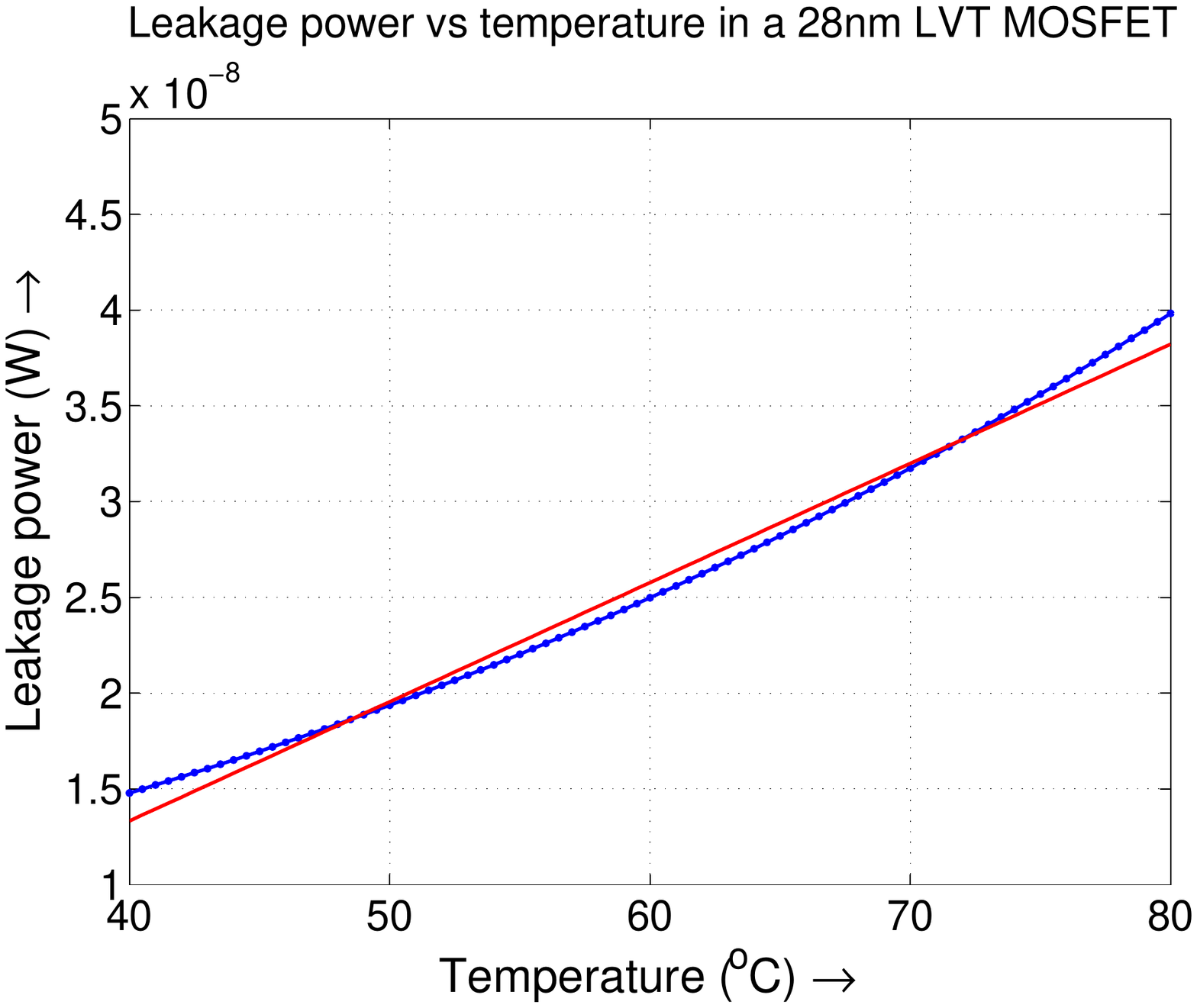}
\caption{Leakage power vs temperature for a 28 nm LVT MOSFET\label{fig:28lvtPT}}
\end{figure}

\begin{figure}[!htbp]
\centering
\includegraphics[scale = .4,trim = {0cm 6cm 0cm 6cm}, clip=true]{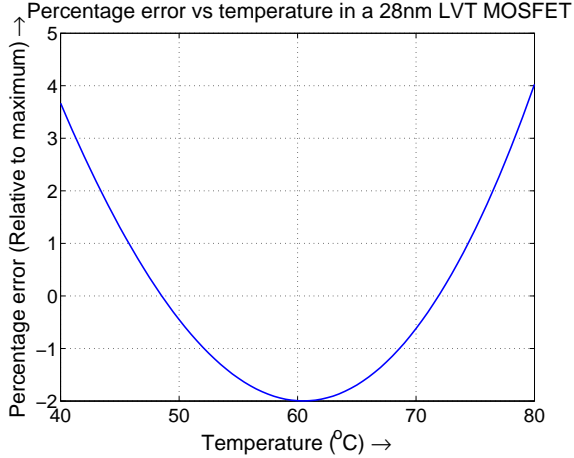}
\caption{Percentage error with a linear leakage model for a 28 nm LVT MOSFET\label{fig:lvterror}}
\end{figure}

We computed the leakage versus temperature characteristics of a 28nm HVT and LVT UMC MOSFET using SPICE simulations. `
Figures~\ref{fig:28hvtPT} to \ref{fig:lvterror} show the leakage-temperature characteristics and the percent error obtained using a linear model (regression line).

\begin{figure}[!htbp]
\centering
\includegraphics[scale = .4,trim = {0cm 6cm 0cm 6cm}, clip=true]{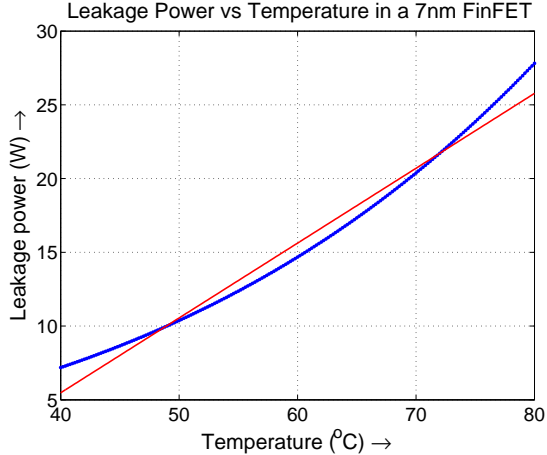}
\caption{Leakage power vs temperature for a 7 nm FinFET\label{fig:7pt}}
\end{figure}

\begin{figure}[!htbp]
\centering
\includegraphics[scale = .4,trim = {0cm 6cm 0cm 6cm}, clip=true]{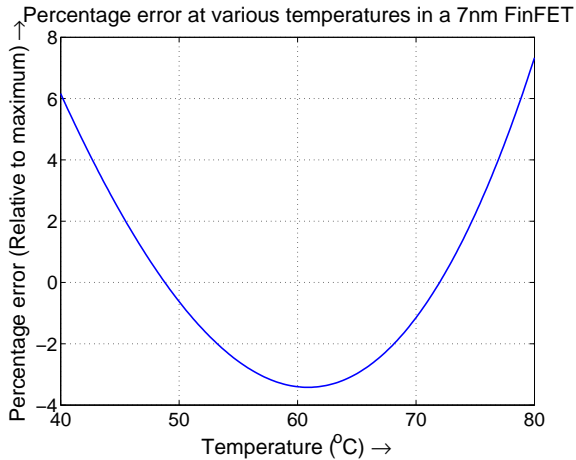}
\caption{Percentage error with linear leakage for a 7 nm FinFET\label{fig:7error}}
\end{figure}

We also simulated a 7 nm multifin device using predictive models \cite{ptm}. The leakage versus temperature characteristics and the percentage error with a linear leakage model is given by Figures~\ref{fig:7pt} and \ref{fig:7error}.

\section{A Review of Leakage Related Papers}

\subsection{Linear Models}

Liu et al.\cite{liu} have linearized the BSIM leakage current equation around a reference temperature $T_{ref}$ using the following equation:
\begin{equation}
\begin{split}
\label{eqn:linIleak}
 I_{leak} = &I_{gate} +  A_s \frac{W}{L}{(\frac{k}{q})}^2 * e^{\frac{q(V_{GS}-V_{th})}{\eta kT_{ref}}} \\&\times \left({T_{ref}}^2 + \left(2T_{ref} - \frac{q(V_{GS} - V_{th})}{\eta k}\right)(T-T_{ref})\right)\\
\end{split}
\end{equation}
They have compared their results with two circuits: a combinational circuit benchmark C7552 and an SRAM, using HSPICE simulations for a 65 nm technology. 
The authors conclude that the linear model is sufficiently accurate within the operating range of ICs. To further improve the accuracy, they use a piecewise linear model. For three or more segments in the piece wise linear model, they found the maximum error to be less than 0.69\%. These results show that a piecewise linear model with a small number of segments is highly accurate. 

\begin{figure}[!htb]
\centering
\includegraphics[scale = 0.4,trim = {0cm 6cm 0cm 6cm}, clip=true]{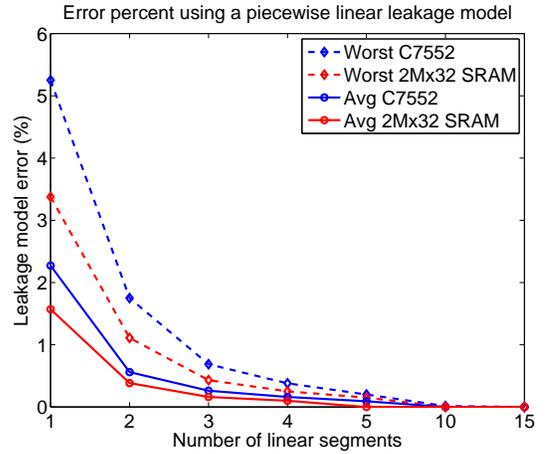}
\caption{Leakage estimation error as a function of the number of segments (adapted from [\cite{liu}])\label{fig:liu2}}
\end{figure}

\subsection{Exponential Models}
\begin{enumerate}

\item 

Martinez et al. \cite{martinez} use an IR camera to obtain thermal images and then map the power to temperature.
They use the following equation to model leakage power:
\begin{equation}
\label{eqn:renau}
P_{leak} = c_0T^2e^{(\frac{c_1}{T})}(1-e^{\frac{c_2}{T}})
\end{equation}
To determine the constants $c_0$, $c_1$ and $c_2$, as well as the dynamic power, the authors use genetic algorithms.

\item 

Singla et al. \cite{singla} have simplified the leakage current using the following equation:
\begin{equation}
\label{eqn:leaksingla}
I_{leak} = c_0T^2e^{(\frac{c_1}{T})} + I_{gate}
\end{equation}

To find the constants, the authors kept their platform in a furnace. The authors ran a light workload on the big core with fixed $f$ and $V_{DD}$ such that the dynamic power did not increase the temperature. They varied the temperature profile from 40\celsius to 80\celsius in steps of 10\celsius. 
These power measurements were used to find the constants and model the leakage power. The leakage power obtained  and the error percent with a linear leakage model is shown in Figure~\ref{fig:singlapt} and \ref{fig:singlaerror}.

\begin{figure}[!htbp]
\centering
\includegraphics[scale = .4,trim = {0cm 6cm 0cm 6cm}, clip=true]{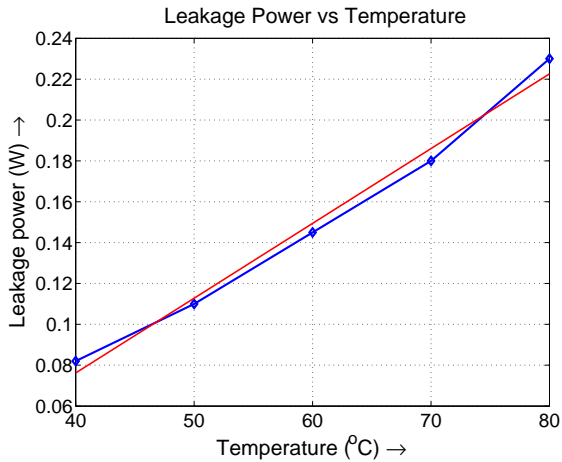}
\caption{Leakage power vs temperature (adapted from \cite{singla}) \label{fig:singlapt}}
\end{figure}
 
\begin{figure}[!htbp]
\centering
\includegraphics[scale = .4,trim = {0cm 6cm 0cm 6cm}, clip=true]{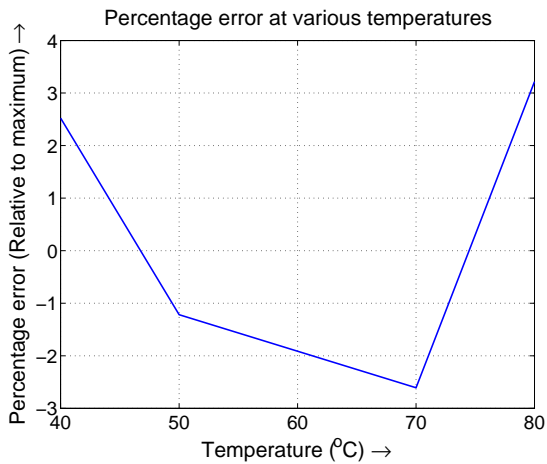}
\caption{Percentage error with a linear leakage model (calculated on the basis of results in \cite{singla})\label{fig:singlaerror}}
\end{figure}

\end{enumerate}

\subsection{Higher Order Polynomial Models}

\begin{enumerate}
\item 

Biswas et al. \cite{isca} perform HSPICE simulations for an SRAM cell at  32nm technology node to model the leakage power relative to 298K. 
They find that a third-order leakage model keeps the error rate below 0.25\%.

\begin{figure}[!htb]
\centering
\includegraphics[scale = 0.4,trim = {0cm 7cm 0cm 6.5cm}, clip=true]{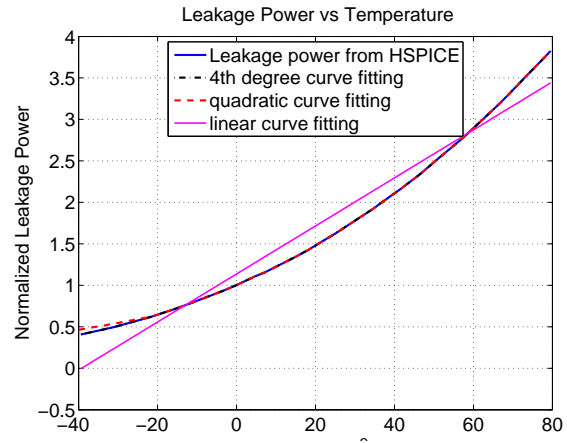}
\caption{Leakage power vs temperature (adapted from \cite{isca})\label{fig:iscafire}}
\end{figure}

\begin{figure}[!htbp]
\centering
\includegraphics[scale = .4,trim = {0cm 6cm 0cm 6cm}, clip=true]{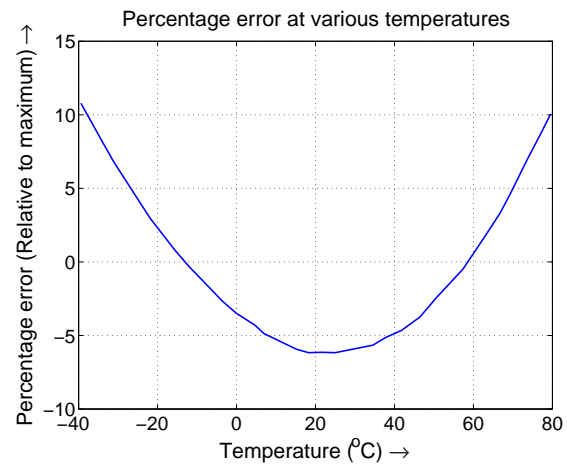}
\caption{Percentage error with a linear leakage model (calculated on the basis of results in \cite{isca})\label{fig:iscaerror}}
\end{figure}

\end{enumerate}

\subsection{Recent Methods: Variation of linear models}
\begin{enumerate}
\item

Wan et al. \cite{wan} consider subthreshold leakage as a linear function of
temperature around a local reference temperature. The reference temperature is updated
when the node temperature differs from the reference temperature by a fixed value
(taken as 10\celsius). When the authors use a single reference temperature for the entire range, they report a maximum error of 1.34\celsius in the temperature calculation, and an error of 3.9\% for power calculation. 

\begin{figure}[!htb]
\centering
\includegraphics[scale = 0.4,trim = {0cm 6cm 0cm 6.5cm}, clip=true]{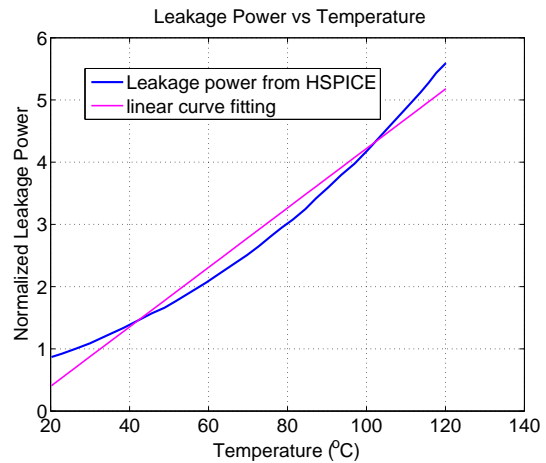}
\caption{Leakage power vs temperature (adapted from \cite{wan})\label{fig:lastpt}}
\end{figure}

\begin{figure}[!htbp]
\centering
\includegraphics[scale = .4,trim = {0cm 6cm 0cm 6cm}, clip=true]{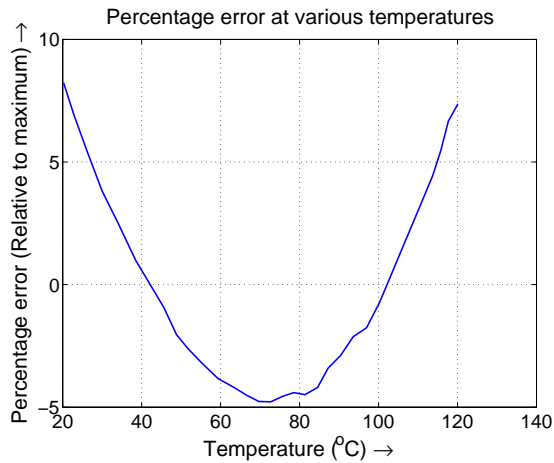}
\caption{Percentage error with a linear leakage model (calculated on the basis of results in \cite{wan})\label{fig:lasterror}}
\end{figure}

\item 

Wang et al. \cite{wang} linearize leakage using multiple expansion points using an equation similar to Equation~\ref{eqn:linIleak}.
To compare the accuracy of using multiple expansion points, they compare their results with TILTS, in which they have implemented a linear leakage model with a single expansion point. The maximum error obtained in temperature estimation using TILTS with linear leakage was 6.45 \celsius with an average error of 0.83 \celsius. Their modified model gave a maximum error of 3.96 K, with average error of 0.17 K.

\item 

Yan et al. \cite{yan} propose a leakage aware analytical thermal modeling
technique for 3D chips assuming a linear leakage model. Their leakage model uses a different expansion
point for each grid in the model. A coarse thermal simulation is first performed
to estimate the expansion points. To further improve the accuracy, they introduce an
additional term for higher order coefficients of leakage power. This term is determined
iteratively.
The authors show that using a uniform linear leakage model, an error of upto
12.67K can be obtained. The authors also show that even the linear model with multiple
expansion points has large errors, while their corrected linearized model is very
accurate.


\end{enumerate}
\section{Timing Analysis of the Discussed Approaches}
We have implemented the techniques discussed in this paper to evaluate the time complexity of these techniques. Table~\ref{tab:speed} lists the time taken to run the mentioned technique using Matlab on an Intel i7 Desktop running Windows 8. 

\begin{table}[!htb]
\small
 \begin{center}
\caption{ Computation time for various models\label{tab:speed}} \begin{tabular}{|p{5cm}|l|} 
\hline
Model & Computation Time (us) \\
\hline
Linear around a reference temperature (Equation \ref{eqn:linIleak})& 19.53\\
\hline
Exponential model 1 (Equation~\ref{eqn:leaksingla}) & 39.08\\
\hline
Exponential model 2 (Equation~\ref{eqn:renau})& 52.34 \\
\hline
Biswas et al. \cite{isca} quadratic model & 20.93\\
\hline
Biswas et al. \cite{isca} cubic model & 25.51\\
\hline
Biswas et al. \cite{isca} fourth degree model & 80.04\\
\hline
\end{tabular}
\end{center}
\end{table}

As we can see from Table~\ref{tab:speed}, the time complexity varies greatly based on the type of leakage model used.
 The difference between the complexity of a linear and a piecewise linear model is not much. However, when a simple exponential model is used (like that in Equation~\ref{eqn:leaksingla}), the time complexity becomes over twice that of a linear model. Also, the error rate using a simple linear model is less than 6\% in most of the cases, which might be sufficient for the early stages of the design process. As stated by Liu et al. \cite{liu}, using a piecewise linear model with just three segments can bring down the error rate to under 1\%. Biswas et al. \cite{isca} state that a third order leakage model keeps the error rate below 0.25\%. However, our simulation results show that this comes at the cost of a 36\% increase in time complexity. We found the error using a quadratic model to be under 1\%, with a 12\% increase in time complexity. Hence we believe that a piecewise linear or a quadratic model should suffice even for high accuracy applications.

\section{Conclusion}
In this paper we have discussed the various leakage power estimation techniques used in the Architecture community, and their time complexity and accuracy. We have concluded that a simple linear model provides an accuracy of over 94\% for most cases, and may suffice in the early stages of the design process. For further improving the accuracy, a piecewise linear model may be used, which brings the accuracy to over 99\%. A quadratic model provides an an accuracy of over 99.3\%, with a 12\% increase in time complexity, which should be sufficient for almost all applications. A simple exponential model increases the time complexity by over 100\% and seems to be an overkill.


\end{document}